\documentclass[twocolumn,aps,prb,floatfix,showpacs]{revtex4}
\usepackage{bm}
\usepackage{amssymb,amsmath}
\usepackage{dcolumn}
\usepackage{graphicx}
\usepackage{bbold}
\usepackage{booktabs}
\usepackage{color}
\usepackage[colorlinks,bookmarks=false,citecolor=blue,linkcolor=blue,urlcolor=blue]{hyperref}

\def\x{\hat{x}}
\def\y{\hat{y}}

\def\calP{{\cal P}}
\def\calB{{\cal B}}
\def\r{{\bf r}}
\def\k{{\bf k}}
\def\S{{\bf S}}
\def\R{{\bf R}}
\def\Q{{\bf Q}}
\def\calC{{\cal C}}
\def\sbar{\bar{s}}
\def\tbar{\bar{t}}
\def\udel{\underline{\delta}}

\begin{document}
\title{ Fourfold degenerate columnar-dimer ground state in square lattice antiferromagnets}
\author{Rakesh Kumar}
\author{Brijesh Kumar} 
\email{bkumar@mail.jnu.ac.in}
\affiliation{School of Physical Sciences, Jawaharlal Nehru University, New Delhi 110067, India}
\date{\today}
\begin{abstract}
We construct and study two frustrated quantum spin-1/2 models on square lattice, which are like the antiferromagnetic  $J_1$-$J_2$ model with some additional four-spin exchange interactions. 
These models admit an exactly solvable case in which the ground state consists of four degenerate columnar-dimer singlet (CDS) configurations. Away from the exact case, we employ bond-operator mean-field theory to investigate the evolution of the ground state by varying the interaction parameters. The mean-field calculation reveals a quantum phase diagram in which the CDS phase undergoes a continuous phase transition to either N\'eel or collinear ordered antiferromagnetic phases.
\end{abstract}

\pacs{75.10.Jm, 75.30.Kz, 75.50.Ee, 75.40.Mg}

\maketitle
\section{Introduction}
The low-dimensional quantum antiferromagnets are exciting physical systems in which  the quantum mechanics, combined with reduced spatial dimensionality and frustration, manifests in a variety of novel properties such as the zero magnetic moment and the loss of long range order  in a {\em spin-liquid} ground state, the fractionalized elementary excitations, namely spinons, or the spontaneous dimerization of spins in the valence bond crystal (VBC) ground states~\cite{Lhuillier, Indrani}.  Over the last two decades, the question of spin-liquid states and fractionalized excitations in two dimensions (2D) has been greatly motivated by the high-$T_c$ superconductivity in the cuprates. Starting with Anderson's proposal of the resonating valence bond theory of superconductivity\cite{RVB_pwa}, a large body of theoretical investigations is centered around the notion that the superconducting state in these materials is derived from a spin-liquid state by  doping it with charge carriers. This novel idea continues to inspire the search for new quantum spin models with novel ground states on two dimensional lattices, and newer methods for studying these models. 

Even prior to the motivations from the cuprates, there has always been a great interest in the studies of quantum spin models obviously for the reasons of explaining magnetic phenomena in real materials. Purely theoretical motivations have also led to interesting developments in this field. One such notable case is the spontaneous dimerization in the exact, doubly degenerate singlet ground state of the Majumdar-Ghosh model~\cite{MG}. It is a one-dimensional quantum spin-1/2 model with antiferromagnetic exchange interactions between the first neighbors ($J_1$) and the second neighbors ($J_2=J_1/2$). This is the simplest known case of a VBC ground state whose wavefunction is a pure direct product of the pairwise singlets (also called the valence bonds, or dimers) between the nearest-neighbor spins. The term VBC refers to a spatially ordered configuration of the valence bonds. This model has inspired a large number of studies on frustrated quantum antiferromagnets, particularly towards constructing new spin models with exact dimer ground states~\cite{BK_linX,MG_motivate1,bose_mitra,MG_motivate2,MG_motivate3}. For example, the Shastry-Sutherland model~\cite{SS} (of which there is a material realization in SrCu$_2$(BO$_3$)$_2$~~\cite{Kageyama,Miya_Ueda}), the Klein's construction~\cite{Klein}, and the AKLT models~\cite{AKLT,AKLT_2} are  some of the more notable of these subsequent works.

Through other very important developments in the form of Lieb-Schultz-Mattis (LSM) theorem~\cite{LSM} and the Haldane's conjecture~\cite{Hald_Conject}, it became possible to make more general statements on the nature of the ground state of quantum antiferromagnets. The LSM theorem states that the ground state of a spin-1/2 antiferromagnetic chain is unique with gapless excitations, or else it is (at least) doubly degenerate. The Haldane's conjecture, on the other hand, suggests a unique ground state with a  spin-gap for an {\em integer} spin chain, thus differentiating between the integer and the half-integer spins~\cite{proof_Hald}. Subsequently, there have been efforts to generalize these results for different situations~\cite{LSM_highD}. Of particular interest to us is a proposal, based on field theoretic analyses~\cite{Haldane_2D,ReadSach_NPhyB}, suggesting the possibility of a fourfold degenerate quantum disordered ground state in the spin-1/2 quantum antiferromagnets on square lattice. In such a disordered state, the local magnetic moment is zero due to quantum fluctuations, but the lattice translational and rotational symmetries can be broken while forming a degenerate valence bond crystal state~\cite{ReadSach_NPhyB}. 

\begin{figure}
\centering
\includegraphics[width=4cm]{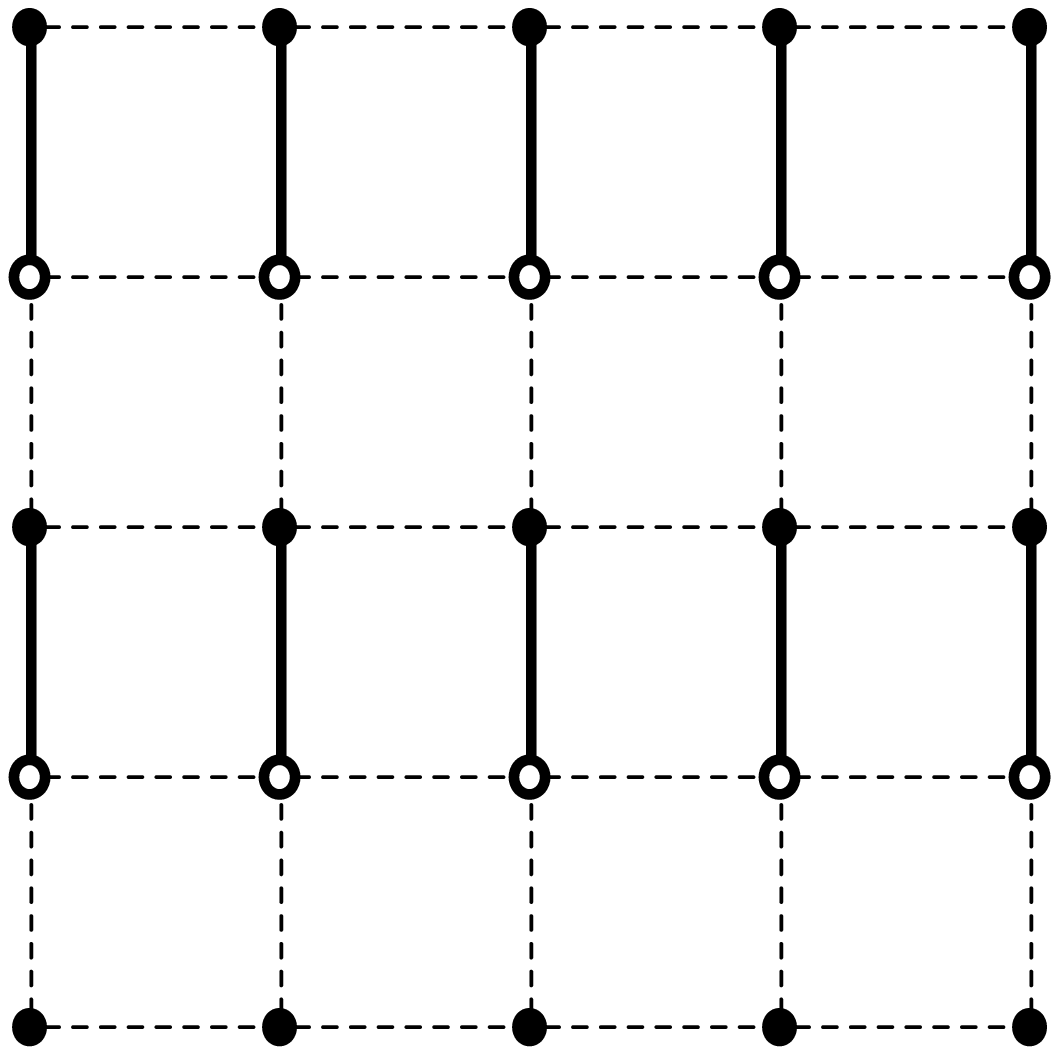} \hfill \includegraphics[width=4cm]{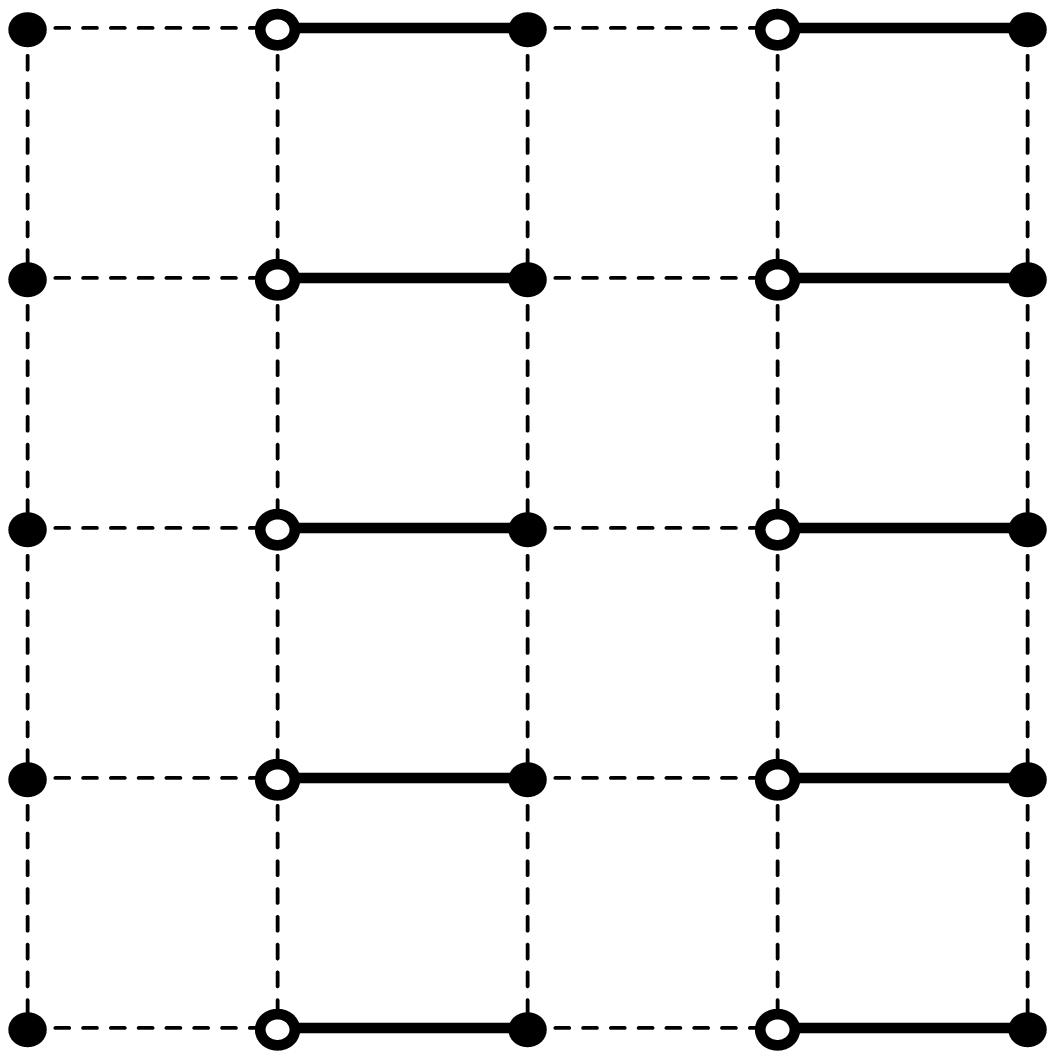}

\vspace{5mm}
\includegraphics[width=4cm]{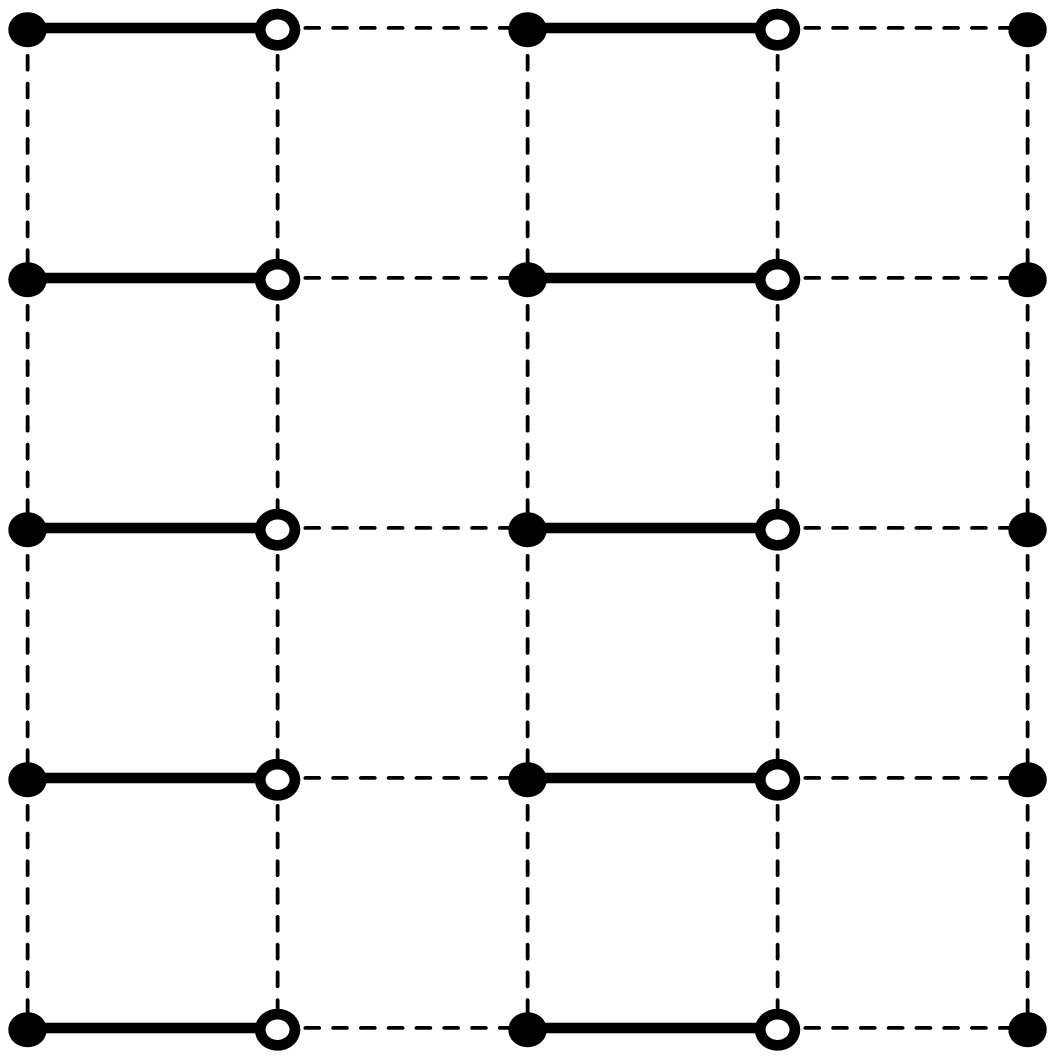} \hfill \includegraphics[width=4cm]{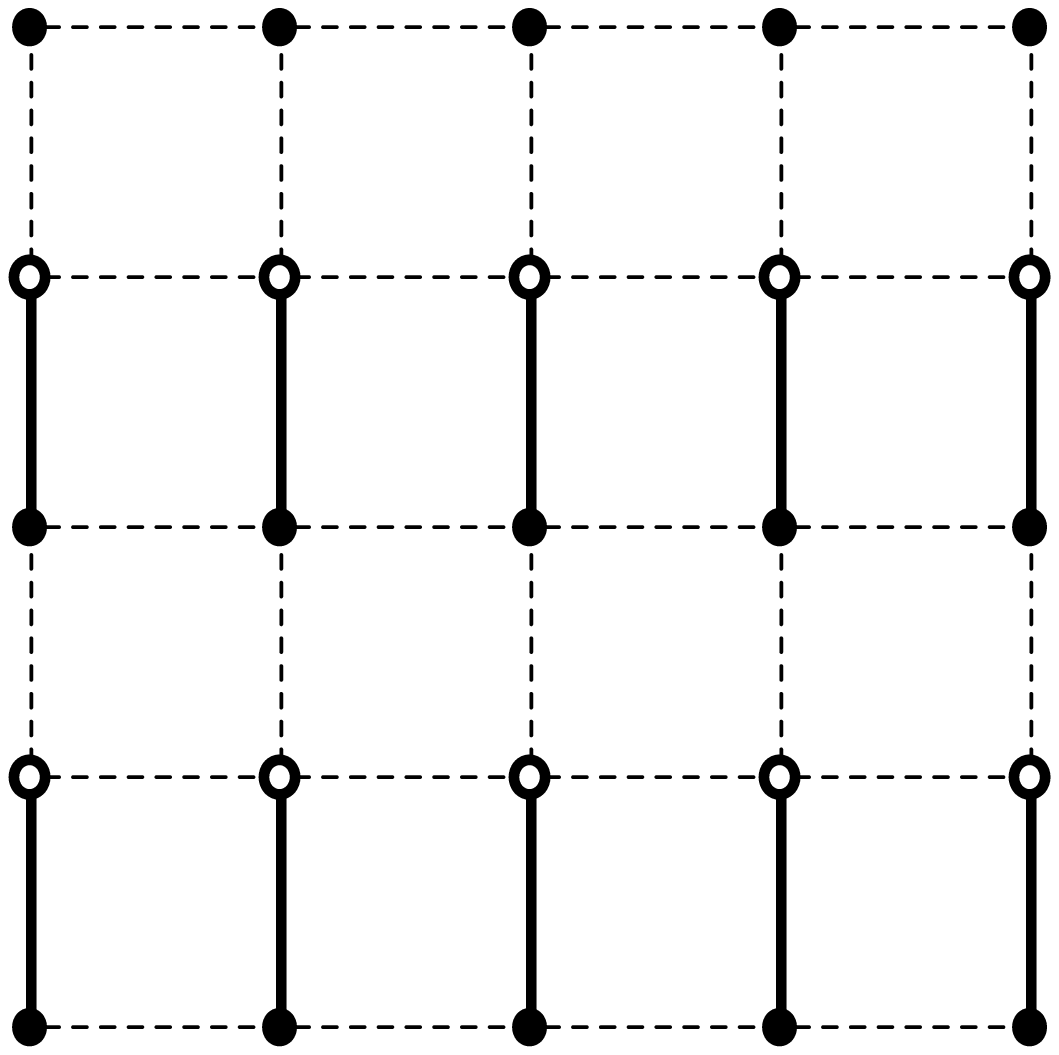}
\caption{
Four columnar-dimer singlet configurations which form the exact ground state of the models $H_{K,1}$ and $H_{K,2}$. A solid line, connecting a filled circle with an empty one, represents a dimer (singlet): \(\frac{1}{\sqrt{2}} |\uparrow\downarrow-\downarrow\uparrow\rangle\). We denote two ends of a singlet bond by a filled and an empty cricle in order to emphasise the antisymmetry of its wavefunction under the exchange of two ends (spins).  }
\label{fig:CDS}
\end{figure}

A popular suggestion for such VBC ground states on square lattice is the columnar-dimer singlet states (CDS) shown in Fig.~\ref{fig:CDS}. Since the ground state of the nearest-neighbor Heisenberg antiferromagnet on square lattice has N\'eel order~\cite{2DHeis}, the CDS phase is expected to emerge only when the ordered state is sufficiently frustrated by adding suitable competing exchange interactions. The nature of the quantum phase transition from an ordered antiferromagnetic (AF) phase to a VBC phase is also a subject of great current interest. Recently, an interesting scenario has been proposed in which an ordered AF phase undergoes a direct second order transition to a VBC phase, through a common quantum critical point~\cite{Senthil1}. In the Landau-Ginzburg approach, the two phases in the ground state can be independently characterized by the non-zero values of the respective order parameters, and the fluctuations thereof. While undergoing a phase transition, it therefore seems very unlikely that the two order parameters will {\em conspire} to continuously vanish exactly at the same point. In the new scenario, however, such a quantum phase transition is described in terms of the spinons, which are `deconfined' at the quantum critical point from which the AF and the VBC phases are claimed to emerge as the interaction parameters vary~\cite{Senthil2}. 

Evidently, it is desirable to find a spin-1/2 quantum antiferromagnet on square lattice which has a fourfold degenerate CDS ground state, and where the scenario of deconfined quantum criticality can possibly be realized. The antiferromagnetic $J_1$-$J_2$ model on square lattice has been studied vigorously by various means for exploring these possibilities (as well as for independent reasons). There was also suggested earlier a certain model with interactions up to fifth neighbor spins, for which the four CDS states were found to be exact eigenstates, but did {\em not} form the exact ground state~\cite{bose_mitra}. The $J_1$-$J_2$ model has N\'eel order in the ground state  when $J_1$ is sufficiently stronger compared to $J_2$, and collinear order (with ordering wavevectors $(\pi,0)$ or $(0,\pi)$) in the opposite limit. Similar to the Majumdar-Ghosh chain, the $J_1$-$J_2$ model on square lattice may be expected to have a  spontaneously dimerized ground state around $J_2\sim 0.5 J_1$. A number of studies have indeed shown a spin gapped quantum disordered phase in the intermediate range of the coupling, $0.4\lesssim J_2/J_1\lesssim0.6$. This intermediate phase is suggested to possess the columnar-dimer order~\cite{vbc1, vbc2, vbc4, vbc5, sach-bhatt, chubokov_2}, but the alternate possibilities (such as a plaquette phase) have also been put forward~\cite{plaquette1, plaquette2, sl2, sl5}. Here, the nature of quantum phase transition from the ordered AF to dimer phase is still under investigation, though there are suggestions of its being a weak first order type~\cite{Kuklov, sirker, Kruger}.

Motivated by these developments, we have constructed new models of spin-1/2 quantum antiferromagnets on square lattice. An important feature of the models presented here is the case when the CDS configurations form an exact fourfold degenerate ground state. This is a case of spontaneous dimerization in the ground state, as the model Hamiltonians respect the symmetries (point group as well translational) of the underlying square lattice. Through these models therefore, we are able to realize a part of our motivation exactly. The models have two parts, one of which is the $J_1$-$J_2$ model, and the remaining part consists of  certain four-spin interactions. For the exact ground state, $J_2=J_1/2$ and the four-spin interactions are important (but not dominant). Therefore, these models can be viewed as two dimensional analogues of the Majumdar-Ghosh chain (in the sense of $J_2/J_1$, and the spontaneous dimerization of nearest-neighbor spins in the ground state of desired degeneracy). Away from the exact case, we formulate a mean-field theory in terms of the bond-operator bosons, to investigate the domain of stability of the CDS phase, and the transition to an ordered phase. In the following sections, we first discuss the models and the exact CDS ground state. Then, we describe the mean-field calculations for the general case, and present the quantum phase diagram. Finally, we conclude with a discussion of results, and a summary.


\section{Models}
\begin{figure}
\centering
\includegraphics[width=4cm]{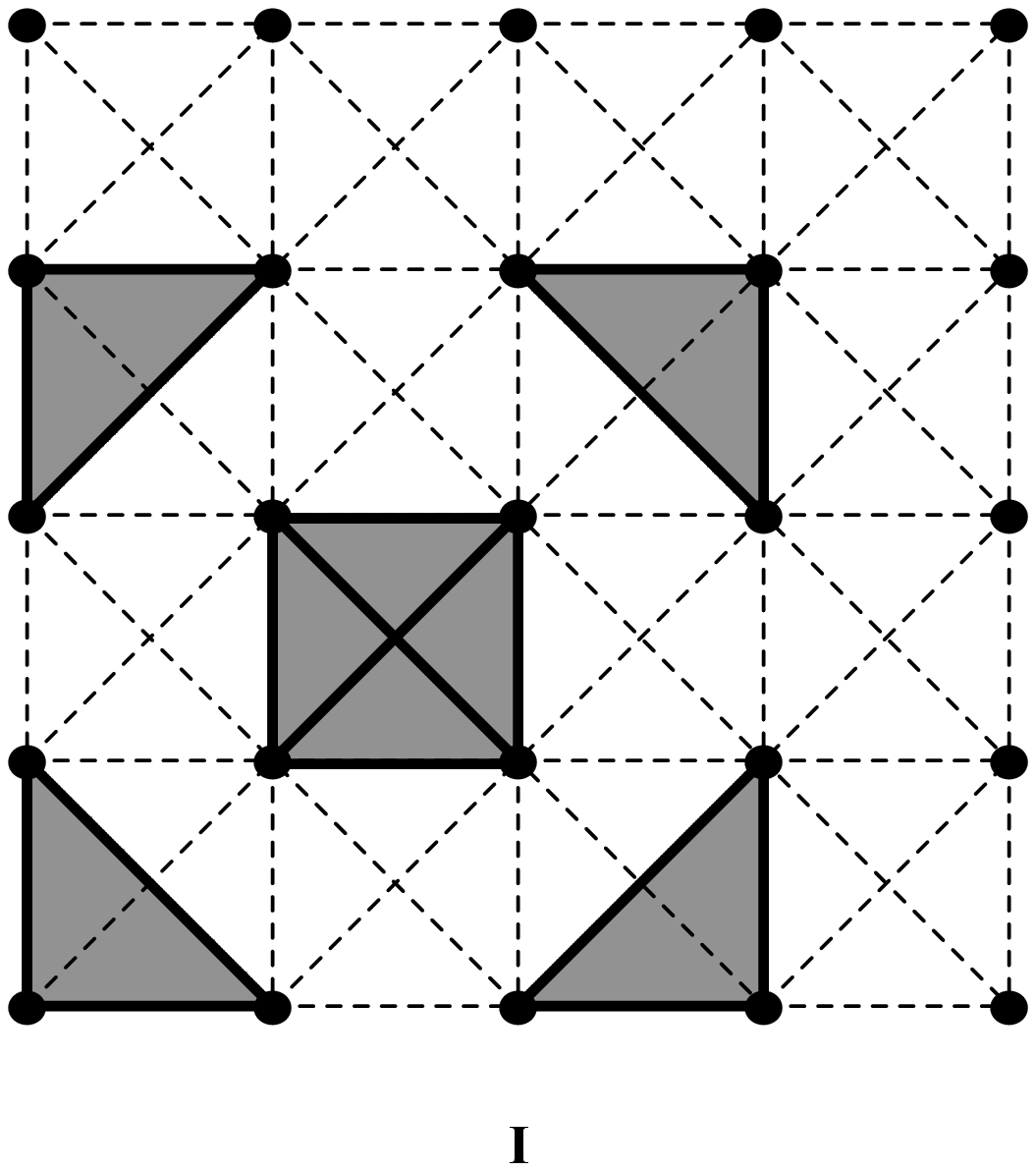} \hfill \includegraphics[width=4cm]{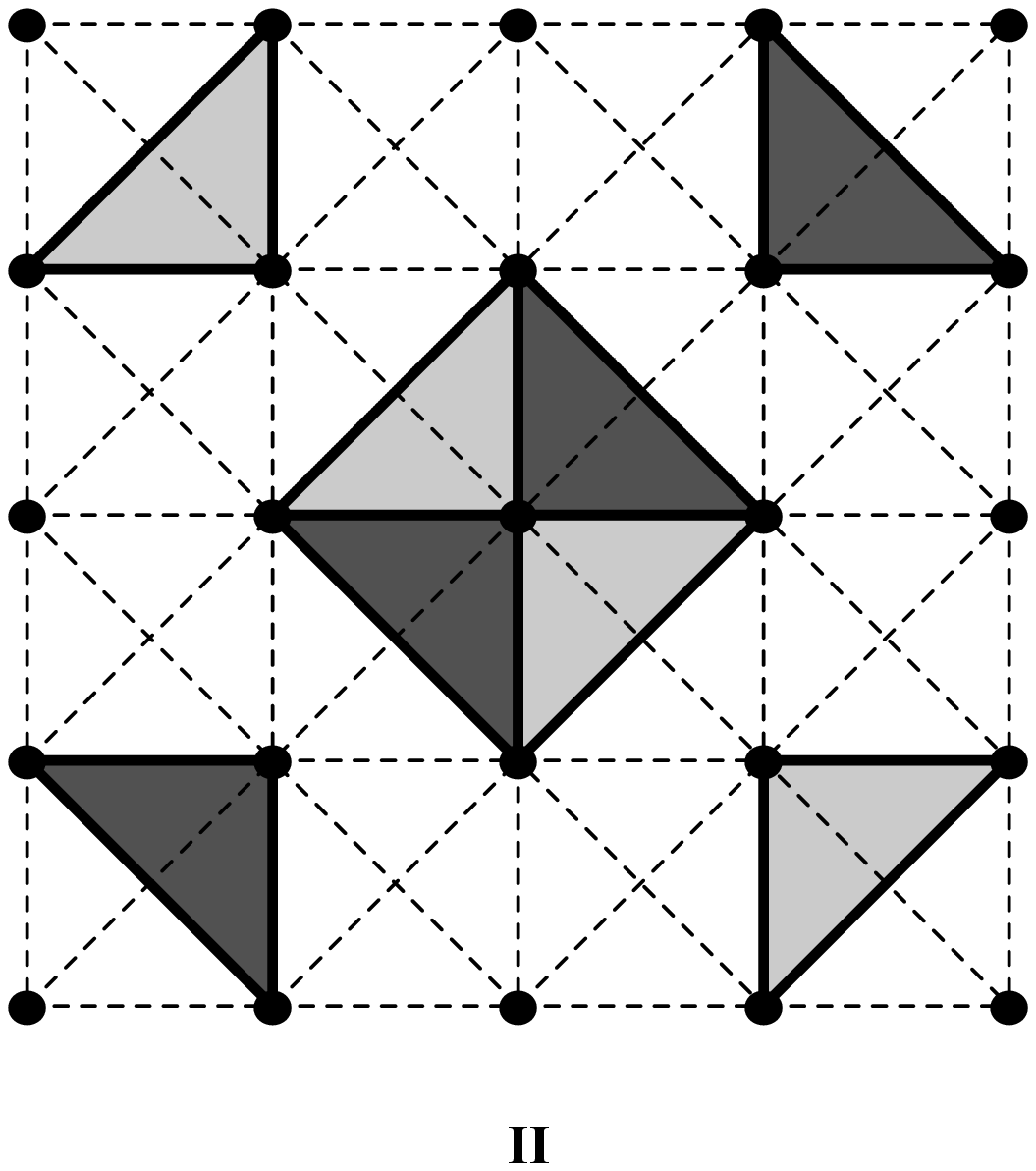}
\caption{Pictures-I and II depict the models $H_1$ and $H_2$, respectively. The dashed background denotes the $J_1$-$J_2$ model. The shaded trianlges denote $\calP_\frac{3}{2}$, and the shaded square (including diagonal lines) stands for $\calB^{(4)}$ (see Eqs.~\ref{eq:p3by2} and~\ref{eq:b4}). The shaded square in picture-I surrounded by four shaded triangles should be understood as a sum of four terms, each of which is a product of the square with one of the triangles. This sum over the whole lattice gives $H_{K,1}$ (Eq.~\ref{eq:HK1}). In picture-II, a pair of shaded triangles with same orientation denotes their product, and together they denote a sum of these four pairwise products. This gives $H_{K,2}$ on square lattice (Eq.~\ref{eq:HK2}).}
\label{fig:models}
\end{figure}

Here, we present two new quantum spin-1/2 models on square lattice. Their scheme of construction involves identifying suitable block-Hamiltonians, and placing these blocks (of spins) appropriately on the lattice such that the CDS configurations are stabilized as ground state. This has been an effective general approach for constructing exactly solvable spin models~\cite{BK_linX}. 
The Hamiltonians of these models are: \( H_1 = H_J + H_{K,1} \) and \(H_2=H_J + H_{K,2}\), in which $H_J$ is the $J_1$-$J_2$ model whose Hamiltonian is written as:
\begin{eqnarray}
H_J &=& J_1\sum_\r \, \left[\S_\r\cdot\S_{\r+a\x} + \S_\r\cdot\S_{\r+a\y}\right] ~+ \label{eq:J1J2} \\
& & \, J_2\sum_\r \, \left[\S_\r\cdot\S_{\r+a\x+a\y} + \S_\r\cdot\S_{\r-a\x+a\y}\right] \nonumber
\end{eqnarray}
where $\r$ is summed over all the sites of square lattice, and $a$ is the lattice constant. We parameterize $J_1$ and $J_2$ as: $J_1=(1-\zeta) J$ and $J_2=\zeta J$ such that $J_1+J_2=J$, where 
$J\ge0$ and $0\le\zeta\le1$. The $H_K$ part of the models is {\em new} and important as it stabilizes the fourfold degenerate CDS ground state. These are written as:
\begin{eqnarray}
H_{K,1} &=& K\sum_\r \sum_{\udel} \, \calP_\frac{3}{2}(\r;\,\udel)~ \calB^{(4)}(\r+\udel;\,\udel)
\label{eq:HK1}\\
H_{K,2} &=& K\sum_\r \sum_{\udel} \, \calP_\frac{3}{2}(\r;\,\udel)~ \calP_\frac{3}{2}(\r+\udel;\,\udel)
\label{eq:HK2}
\end{eqnarray}
where $\udel=(\delta_x, \delta_y)$ with $\delta_x=\pm a\x$ and $\delta_y=\pm a\y$. The coupling $K$ is taken to be positive, and parameterized together with $J$ as: $K=\kappa$ and $J=1-\kappa$ such that $J+K=1$ and $0\le\kappa\le1$. These models are pictorially described in Fig.~\ref{fig:models}. 

Operator $\calP_\frac{3}{2}$ in Eqs.~\ref{eq:HK1} and~\ref{eq:HK2} is a projection operator for a triangular block which annihilates $S_{tot}=1/2$ states, where $S_{tot}$ denotes the total spin of a block. The block operator $\calB^{(4)}$, of four spin-1/2s, is not a projector, but annihilates $S_{tot}=0$ sector, and is positive otherwise. In terms of the spin operators, these are defined as:
\begin{eqnarray}
\calP_{\frac{3}{2}}(\r;\, \udel) &=& \frac{1}{3}\left[\left( \S_\r + \S_{\r+\delta_x} + \S_{\r+\delta_y} \right)^2 - \frac{3}{4} \right] \label{eq:p3by2}\\
&=& \frac{1}{2} + \frac{2}{3}\, h_3(\r;\, \udel) \label{eq:h3}\\
\calB^{(4)}(\r;\, \udel) &=& \frac{1}{6}\left( \S_\r + \S_{\r+\delta_x} + \S_{\r+\delta_y} + \S_{\r+\delta_x + \delta_y} \right)^2 \label{eq:b4}\\
&=& \frac{1}{2} + \frac{1}{3}\, h_4(\r;\, \udel) \label{eq:h4}
\end{eqnarray}
Expanding Eqs.~\ref{eq:p3by2} and~\ref{eq:b4} results in Eqs.~\ref{eq:h3} and~\ref{eq:h4} 
where $h_3(\r;\, \udel)= (\S_\r\cdot\S_{\r+\delta_x} + \S_\r\cdot\S_{\r+\delta_y} +  \S_{\r+\delta_x}\cdot\S_{\r+\delta_y})$, and \(h_4(\r;\, \udel)=( \S_\r\cdot\S_{\r+\delta_x} +\S_\r\cdot\S_{\r+\delta_y} + \S_\r\cdot\S_{\r+\delta_x+\delta_y} + \S_{\r+\delta_x}\cdot\S_{\r+\delta_y} + \S_{\r+\delta_x}\cdot\S_{\r+\delta_x +\delta_y} + \S_{\r+\delta_y}\cdot\S_{\r+\delta_x + \delta_y})\), are the exchange Hamiltonians of the `completely-connected' three- and four-spin blocks respectively~\cite{BK_linX}. Clearly, $\calP_\frac{3}{2}$ is $0$ or $1$ for $S_{tot}=1/2$ or $3/2$ respectively. The eigenvalue of $\calB^{(4)}$ is $0$ when the spins of a square plaquette form a singlet, and it is $1/3$ or $1$ for $S_{tot}=1$ or $2$ respectively. The $H_{K,1}$ therefore describes the interaction of a square block with its neigbhoring triangles, and $H_{K,2}$ is an interaction between triangles. Recently, a model with fourfold degenerate Shastry-Sutherland ground state on square lattice has been constructed using a $\calP_{\frac{3}{2}}$-$\calP_{\frac{3}{2}}$ 
interaction~\cite{4SS}, which has inspired the construction of $H_{K,2}$ in the present work.

Rewriting $H_1$ and $H_2$ in terms of $h_3$ and $h_4$, by putting Eqs.~\ref{eq:h3} and~\ref{eq:h4} into $H_{K,1}$ and $H_{K,2}$, gives us: \( H_{1,~2}= H_0 + V_{1,~2} + K L\), where $H_0$, $V_1$ and $V_{2}$ are defined below (and, $L$ is the total number of lattice sites; also, $H_{1,~2}$ reads as $H_1$ or $H_2$, and similarly for $V_{1,~2}$).
\begin{eqnarray}
H_0 &=& \tilde{J}_1\sum_\r\, \left[\S_\r\cdot\S_{\r+a\x} + \S_\r\cdot\S_{\r+a\y}\right] ~ +\label{eq:H0} \\
& & \tilde{J}_2 \sum_\r \, \left[\S_\r\cdot\S_{\r+a\x+a\y} + \S_\r\cdot\S_{\r-a\x+a\y}\right] \nonumber\\
V_1 &=& \frac{2K}{9} \sum_\r\sum_{\udel}\, h_3(\r;\, \udel)\, h_4(\r+\udel;\, \udel) \label{eq:v1}\\
V_2 &=& \frac{4K}{9} \sum_\r\sum_{\udel}\, h_3(\r;\, \udel)\,h_3(\r+\udel;\, \udel) \label{eq:v2}
\end{eqnarray}
Here, $\tilde{J}_1= \left[ (1-\zeta)J + \frac{8}{3}K \right]$ and $\tilde{J}_2=\left[ \zeta J+\frac{4}{3}K \right]$ are the effective $J_1$ and $J_2$ exchange interactions.   
Even when the `external' $J$ is zero ($i.e.$, $\kappa=1$), the `actual' interactions, $\tilde{J}_1$ and $\tilde{J}_2$, are non-zero, and interestingly their ratio is exactly $1/2$. Also, in this case, the typical value of the four-spin exchange energy ($|\langle V_{1,~2}\rangle|/L\sim K$) would be less than (or comparable to) that of the two-spin exchange ($|\langle H_0\rangle|/L \sim 2K$)~\cite{fn-1}. In this sense, the four-spin exchange is important, but {\em not} dominant compared to the usual two-spin exchange interaction.

\subsection{Exact ground state}
An important feature of these models is an exact fourfold degenerate columnar-dimer ground state that exists for $\kappa=1$.  The models in this case consist purely of $H_K$ part, for which $\tilde{J}_2/\tilde{J}_1=1/2$, akin to the Majumdar-Ghosh model. For $\kappa<1$, the CDS configurations (see Fig.~\ref{fig:CDS}) do not form the exact eigenstates of $H_1$ and $H_2$. This case will be studied within a mean-field scheme in the next section. Presently, we discuss the exact ground state. Consider $H_{K,1}$ on a square lattice in toroidal geometry with even number of sites. In this model, every square block interacts with its four neighboring triangles (see Fig.~\ref{fig:models}), expressed as a product of $\calB^{(4)}$ with $\calP_\frac{3}{2}$. Since the lowest eigenvalue of $\calB^{(4)}$ and $\calP_\frac{3}{2}$ operators is zero, a `trial' wavefunction $|\psi\rangle$ for the ground state of the model must satisfy the inequality: \( 0 \le E_g \le \langle\psi|H_{K,1}|\psi\rangle \), where $E_g$ is the ground state energy. If we can find a $|\psi\rangle$ for which the upper bound of the inequality equals its lower bound ($i.e.$, zero), then $|\psi\rangle$ is a ground state with $E_g=0$. The same is also true for $H_{K,2}$.

To construct such an eigenstate, identify the `local' spin wavefunctions which are annihilated by the block-interaction, $\calB^{(4)} \otimes \calP_\frac{3}{2}$. This can be achieved by forming a singlet state on a $\calB^{(4)}$ block, or a doublet ($S_{tot}=1/2$) on a triangle. A singlet state on a square block can be formed by making two dimers. Making one dimer on a triangle leaves the third spin `free', thus forming a doublet state. A zero energy eigenstate on the full lattice can therefore be constructed in the following way: make {\em only} those dimers which either sit on a square, or on a triangle; try forming  a configuration of dimers on the square lattice such that every block-interaction term in $H_{K,1}$ is simultaneously satisfied ($i.e.$, becomes zero). We follow this procedure, considering all allowed dimers at the block level, and find the {\em four} columnar-dimer configurations, shown in Fig.~\ref{fig:CDS}. These four CDS states clearly form the exact  ground state of $H_{K,1}$,  because every block-interaction term in $H_{K,1}$ annihilates them. Following the same procedure, we find that the four CDS states also form the exact ground state of  $H_{K,2}$. In this case, the interaction between triangles becomes zero.

The CDS wavefunctions can explicitly be written as:
\begin{eqnarray}
|CDS_{x,1}\rangle &=& \otimes\prod_{l=1}^{L_x/2} \prod_{m=1}^{L_y}
                                         \left[(2l-1,m)\, ,\,(2l,m)\right]\\ 
|CDS_{y,1}\rangle &=& \calC_4 \, |CDS_{x,1}\rangle \\
|CDS_{x,2}\rangle &=& \calC_4^2 \, |CDS_{x,1}\rangle \\
|CDS_{y,2}\rangle &=& \calC_4^3 \, |CDS_{x,1}\rangle 
\end{eqnarray}
where $[(n_1,m_1),(n_2,m_2)]$ denotes a singlet between two spins with coordinates $(n_1,m_1)$ and $(n_2,m_2)$ on square lattice ($L_x=L_y$, and $L=L_x L_y$)~\cite{fn-2}, and $\calC_4$ denotes the rotation by $\pi/2$. The subscripts $x$ or $y$ in the state-labels denote whether the dimers are formed along $\x$ or $\y$-directions, and $1$ or $2$ denote two states of $x$ or $y$ type. For reference, we can take the bottom-left picture in Fig.~\ref{fig:CDS} as $|CDS_{x,1}\rangle$. Clearly, the CDS states break the rotational symmetry, and also the lattice translation, as $|CDS_{x,1}\rangle$ transforms to $|CDS_{x,2}\rangle$ under the translation along $\x$-direction and $|CDS_{y,1}\rangle$ transforms to $|CDS_{y,2}\rangle$ under the translation along $\y$-direction. Also, the CDS ground state has a gap to the spin  excitations. The energy gap to a localized triplet excitation, created by breaking a singlet bond, is estimated to be $8/3$ (for $\kappa=1$). For $\kappa<1$, the triplet excitation is expected to hop, and therefore,  the spin-gap will be reduced.

In principle, one could envisage the possibility of a `fifth' state in the exact ground state. To show rigorously that the four CDS configurations are the {\em only} states in the exact ground state is a hard problem. However, as stated above, the explicit construction gives only four  CDS configurations in the dimer ground state. We support it by the following counting argument~\cite{4SS}. On a lattice of $L$ sites, there are exactly $L/2$ dimers in any valence bond state. Since there are exactly $4L$ number of block-interaction terms (both in $H_{K,1}$ and $H_{K,2}$), therefore on the average, a dimer is to be shared between eight block-interaction terms. There are only two types of allowed dimers: 1) between first-neighbors, and 2) between second-neighbors. A second-neighbor dimer is common only to four block-interaction terms, whereas a first-neighbor dimer is shared exactly between eight block-interaction terms. Therefore, a dimer eigenstate of the models must be made of nearest-neighbor dimers only. Also, it is necessary that only one of the two interacting blocks becomes zero, otherwise at least one other block-interaction term will remain unsatisfied~\cite{fn-3}.  Furthermore, there are four ways in which a given spin can form a dimer with its nearest-neighbors (two each along $x$- and $y$-axes). Now, if we construct a dimer ground state observing all these constraints, we only generate four CDS configurations (one each for  the four choices of the initial dimer). This simple argument also demonstrates the dependence of the degeneracy of a dimer ground state on the (relevant) coordination of a spin~\cite{fn-4}. For example, the fourfold degeneracy of the Shastry-Sutherland ground state in a recent model~\cite{4SS} is consistent with the second-neighbor coordination on square lattice being four.
 \section{Triplon mean-field theory}
Now, we study the general case of $H_1$ and $H_2$ in which $\kappa$ and $\zeta$ take arbitrary values in the interval $[0,1]$. As mentioned earlier, the spin-gapped CDS state is no more the exact eigenstate of $H_{1,~2}$ for $\kappa <1$. However, we expect that the spontaneous dimerization in the ground state will continue to persist until the spin-gap is closed. We therefore want to investigate, within some approximate but physically acceptable scheme, the domain of existence of the CDS phase in the $\kappa$-$\zeta$ plane. For this purpose, we choose to work with bond-operator representation of spin-1/2 operators\cite{sach-bhatt,chub}, which provides an appropriate tool for studying a spin system with dimerization. We formulate a simple mean-field theory of $H_{1,~2}$ in terms of the bond-operators, and calculate the quantum phase diagram within this approach. An important feature of this formulation is that we don't have to guess which antiferromagnetically ordered phase will the CDS phase go into. It is decided simply by the wavevector at which the spin-gap closes, and the {\em triplons} condense. A triplon is a triplet quasi-particle residing on a bond. We will formulate this mean-field theory in the present section, and the results of the self-consistent calculation will be discussed in the next section.
 \subsection{Bond-Operator Representation}
The Hilbert space of a pair of $S=1/2$ spins, say $\S_{1}$ and $\S_{2}$, is spanned by a singlet state $|s\rangle$ and three triplet states $|t_{x}\rangle$, $|t_{y}\rangle$ and $|t_{z}\rangle$. In the bond-operator representation, these spin states are created out of a vacuum $|0\rangle$ by means of a `singlet'  and three `triplet' creation operators, as defined below.
\begin{eqnarray}
|{ s} \rangle &=& \frac{1}{\sqrt{2}}| \uparrow \downarrow  - \downarrow \uparrow \rangle 
:= s^{\dag} |0 \rangle \label{eq:s} \\
|{ t}_x \rangle &=& \frac{-1}{\sqrt{2}}| \uparrow \uparrow - \downarrow \downarrow\rangle 
:= { t}_x^{\dag} |0 \rangle \label{eq:tx}\\
|{ t}_y\rangle &=& \frac{i}{\sqrt{2}}| \uparrow \uparrow +  \downarrow \downarrow \rangle   
:= { t}_y^{\dag} |0 \rangle \label{eq:ty}\\
|{ t}_z\rangle &=& \frac{1}{\sqrt{2}}| \uparrow \downarrow + \downarrow \uparrow \rangle 
:= { t}_z^{\dag} |0 \rangle \label{eq:tz}
\end{eqnarray}
The bond-operators $s^\dag$ and $t_\alpha^\dag$ ($\alpha=x, y, z$) are canonical bosons with a {\em constraint}: $s^\dag s + t_\alpha^\dag t_\alpha=1$, where repeated Greek index is summed over. In terms of these bosons, the spin operators, $\S_1$ and $\S_2$,  can be expressed as:
\begin{eqnarray}
S_{1\alpha} &=& \frac{1}{2} \left(s^\dag t_\alpha + t_\alpha^\dag s - 
i\epsilon_{\alpha\beta\gamma} t_\beta^\dag t_\gamma \right) \label{eq:S1_BO}\\
S_{2\alpha} &=& \frac{1}{2} \left(-s^\dag t_\alpha - t_\alpha^\dag s - 
i\epsilon_{\alpha\beta\gamma} t_\beta^\dag t_\gamma \right) \label{eq:S2_BO}
\end{eqnarray}
where $\epsilon_{\alpha\beta\gamma}$ is the totally antisymmetric tensor. Eqs.~\ref{eq:S1_BO} and~\ref{eq:S2_BO} satisfy the spin algebra (by making use of the boson commutation relations and the constraint), and hence define the bond-operator representation\cite{sach-bhatt} of two $S=1/2$ spins. And, $\S_1\cdot\S_2=-\frac{3}{4}s^\dag s+ \frac{1}{4}t^\dag_\alpha t_\alpha$.

In principle, one could just rewrite any spin-1/2 model in terms of the bond-operators. But this alone does not help, because now we have a constrained interacting boson problem. However, within a physically meaningful approximation, this representation can still be used for studying interesting problems where we have a scope for dimerized singlet ground states. The physical setting, in qualitative terms, of such a problem is simple: we have a background of dimers with some {\em mean} singlet amplitude per bond, in which an elementary triplet excitation can be created by breaking a singlet bond; this triplet, in general, can disperse through the background of singlets, assisted by the exchange interactions present in the system. We call this dispersing {\em bond}-triplet a {\em triplon} (in contrast to a magnon, which is also a spin-1 excitation, but created by a single spin-flip). In terms of the bond-operators, this implies a simple mean-field theory, in which we replace $s$ and $s^\dag$ by a $c$-number $\sbar$, which quantifies mean singlet amplitude on every dimer. The resultant Hamiltonian is a model of interacting triplons, which can further be simplified by ignoring the interaction (similar to the spin-wave analysis of a magnetically ordered ground state). This {\em two-step} simplification results in a problem which can be exactly diagonalized. On each dimer {\em locally}, the bosons must also satisfy the constraint. This is however done, on average, by introducing a {\em global} chemical potential. Besides the dimerized ground states, we can also describe magnetically ordered phases in the same formulation, by the condensation of triplons. Therefore, in this mean-field approximation, we can also study  quantum phase transitions in the ground state, by varying suitable parameters in the problem.
\subsection{Mean-Field Theory}
We now derive the mean-field triplon Hamiltonian for $H_{1,~2}$.  
Since there are four CDS configurations in the exact ground state, we can choose any one of these as a reference state for bond-operators. We take $|CDS_{x,1}\rangle$ as the reference state, and represent the spin models in terms of the bond-operators. Though we would like to work with all four  CDS states at the same time, but we can not. This is a limitation of the present approach. It will, nevertheless, give us a good idea of what happens to the CDS phase by varying $\kappa$ and $\zeta$. 

\begin{figure}
\centering
\includegraphics[width=6cm]{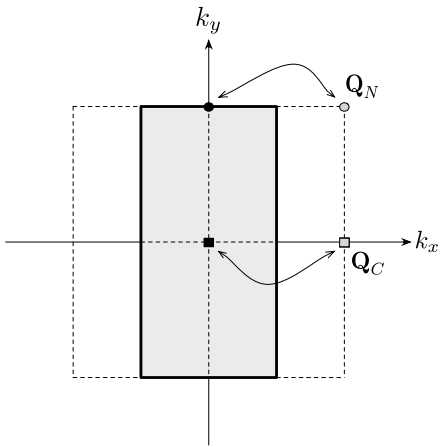}
\caption{The shaded rectangular region denotes the Brilluoin zone of the CDS state (CDS-BZ), while the dashed-outlined big square is the Brillouin zone of the underlying square lattice (Sqr-BZ). The N\'eel ordering wavevector $\Q_N=(\pi,\pi)$ in Sqr-BZ maps to $(0,\pi)$ in CDS-BZ, and similarly, the collinear ordering wavevector $\Q_C=(\pi,0)$ maps to $(0,0)$.} 
\label{fig:CDS_BZ}
\end{figure}

We substitute Eqs.~\ref{eq:S1_BO} and~\ref{eq:S2_BO} for the spin operators in $H_0$, $V_1$ and $V_2$, and incorporate the constraint on bond-operators by including an extra term: $\sum_\R \mu_\R(1-s_\R^\dag s^{ _{ }}_\R -t^\dag_{\R\alpha} t^{ _{ }}_{\R\alpha})$, in the Hamiltonian (here, $\R$ denotes the position of a dimer). The Lagrange multipliers are taken to be uniform, that is $\mu_\R =\mu$. By making the simplifying approximations discussed above, and Fourier transforming the triplon operators, we get the following trilpon Hamiltonian for both $H_1$ and $H_2$.
\begin{eqnarray}
H_t &=& E_0 + \frac{1}{2}\sum_{\k} \left\{ \left(\lambda - \sbar^2\xi_\k\right)\left[ t^\dag_{\k\alpha}t^{ }_{\k\alpha} + t^{ }_{-\k\alpha}t^\dag_{-\k\alpha}\right] \right. \nonumber \\ 
&& \hspace{18mm}\left. - \sbar^2 \xi_\k \left[ t^\dag_{\k\alpha}t^\dag_{-\k\alpha} +  t^{ }_{-\k\alpha}t^{ }_{\k\alpha}\right]\right\}
\label{eq:Ht}
\end{eqnarray}
where
\begin{equation}
\lambda=\frac{1}{4}(1-\zeta)J+ \frac{2}{3}K- \mu
\label{eq:lambda_effective}
\end{equation}
is the effective chemical potential of triplons, 
\begin{eqnarray}
E_0&=&\frac{L}{2}\left\{ \frac{(1-\zeta)}{4}J + \frac{8K}{3}-\frac{5\lambda}{2} 
+ \lambda\sbar^2 \right. \nonumber \\
 && \hspace{5mm}\left.- \bar{s}^2\left[(1-\zeta)J+\frac{8K}{3}\right] \right\} \, \equiv e_0 L
 \label{eq:E0}
\end{eqnarray}
is a constant term, and $\xi_k$ is written as: 
\begin{eqnarray}
\xi_\k&=&J\left\{ (1-\zeta)\left(\cos{2k_x}-2\cos{k_y}\right)/2 ~+  \right. \nonumber\\
&& \left. \zeta\left(1+\cos{2k_x}\right)\cos{k_y} \right\} +  \frac{4K}{3}(1-\sbar^2)\left(\cos{2k_x}\right. \nonumber \\
&&\left. -\cos{k_y}+\cos{2k_x}\cos{k_y}\right)
\label{eq:xi_k}
\end{eqnarray}
Remember that $K=\kappa$, $J=1-\kappa$, and $L$ is the total number of lattice sites. Also, the lattice constant $a$ has been put equal to 1. The Fourier transformation of the triplon operators is defined as: \( t_{\R\alpha}=\sqrt{\frac{2}{L}}\sum_\k e^{i\k\cdot\R}t_{\k\alpha} \), where $\k$ belongs to the CDS Brilluoin zone (CDS-BZ) which is folded half-way along $x$-axis with respect to the Brilluoin zone of square lattice (Sqr-BZ) (see Fig.~\ref{fig:CDS_BZ}).
 
We diagonalize $H_t$ by  making Bogoliubov  transformation: $ t_{{\bf k}\alpha}= \gamma^{ }_{\k\alpha}\cosh{\theta_{\k}} + \gamma_{-\k\alpha}^{\dag}\sinh{\theta_{\k}}$, where $\gamma^{ }_{\k\alpha}$ are new bosons. Rewriting $H_t$ in terms of $\gamma^{ }_{\k\alpha}$, and demanding the off-diagonal terms to be absent gives the equation: $\tanh{2\theta_\k}= \sbar^2\xi_\k/(\lambda-\sbar^2\xi_\k)$, for $\theta_\k$. And, we get the following diagonal form of the mean-field Hamiltonian.
 \begin{equation}
H_t = E_{0}+\sum_{\k}\, E_{\k}\left(\gamma^{\dag}_{\k\alpha}\gamma^{ }_{\k\alpha} + \frac{3}{2} \right)
\label{eq:Ht_diagonal}
\end{equation}
Here, $E_{\k}=\sqrt{\lambda(\lambda-2\sbar^2\xi_k)} \ge 0$ is the triplon quasi-particle dispersion. There are three degenerate quasi-particle branches (each corresponding to a different $\alpha$). The vacuum of the $\gamma^{ }_{\k\alpha}$ bosons is the ground state of $H_t$, with ground state energy density, $e_g$:
\begin{equation}
e_g[\lambda,\sbar^2] = e_0 + \frac{3}{2L}\sum_\k\, E_\k \label{eq:eg}
\end{equation}
where $e_0=E_0/L$. Minimizing $e_g$ with respect to $\lambda$ and $\sbar^2$ will give us the mean-field solution. Since we are interested in understanding the competing phases in the ground state, we restrict our investigation to $T=0$. Though, in principle, we could also formulate a finite temperature mean-field theory.
\subsubsection{Gapped CDS Phase}
In the CDS phase, it costs finite energy to create a triplon. This manifests as a gap in the quasi-particle energy, that is $E_\k >0$. The mean-field parameters $\sbar^2$ and $\lambda$ in this ground state are determined by iteratively solving the following self-consistent equations. 
\begin{eqnarray}
\sbar^2&=& \frac{5}{2}-\frac{3}{L}\sum_{\k} \frac{\lambda-\sbar^2\xi_\k}{E_{\k}} \label{eq:sbar_cds}\\
\lambda &=& (1-\zeta)J+\frac{8}{3}K + \frac{3\lambda}{L}\sum_\k \frac{\eta_\k}{E_\k} \label{eq:lambda}
\end{eqnarray}
Here, \( \eta_\k = \xi_\k - \frac{4K}{3}\sbar^2(\cos{2k_x}-\cos{k_y}+\cos{2k_x}\cos{k_y}) \). These self-consistent equations are a saddle point condition on $e_g[\lambda,\sbar^2]$, that is: $\partial e_g/\partial\lambda=0$ and $\partial e_g/\partial\sbar^2=0$, respectively.
As long as the triplon gap is non-zero, the denominators in Eqs.~\ref{eq:sbar_cds} and~\ref{eq:lambda} would not have any problem, and the two equations can be solved self-consistently. 

The $\sbar^2$ thus calculated directly measures the expectation value of the singlet projection operator on a dimer in the mean-field CDS ground state. The singlet projector on a dimer is defined as: $\frac{1}{4}-\S_1\cdot\S_2$, which in terms of the bond-operators is exactly equal to $s^\dag s$, and therefore, has $\sbar^2$ as its mean-field value. In fact, it is encouraging to state that our {\em mean-field calculation  is exact for} $\kappa=1$, in the sense that it gives $\sbar^2=1$, which is same as the expectation value of the singlet projector in the exact CDS ground state.
The mean-field theory also gives {\em zero magnetic moment} in the gapped case, as expected for a quantum disordered state. It can be shown by noting that $\langle S_{1\alpha}\rangle=\sbar\langle t_\alpha^\dag + t^{ _{ }}_\alpha\rangle/2\, -i\epsilon_{\alpha\beta\gamma} \,\langle t^\dag_\beta t^{ _{ }}_\gamma\rangle$ in the mean-field theory, and similarly $\langle S_{2\alpha}\rangle$. Since there is no  mixing of modes with different values of $\alpha$ (see Eqs.~\ref{eq:Ht} and~\ref{eq:Ht_diagonal}), and the ground state is a `normal' boson vacuum ($i.e.$, $\langle t_\alpha\rangle=\langle t^\dag_\alpha\rangle=0$), the local magnetic moments, $\langle \S_1\rangle=\langle\S_2\rangle=0$. Hence, the mean-field ground state is quantum disordered, when triplons are gapped. The actual dependence of the gap and $\sbar^2$ on $\kappa$ and $\zeta$ is computed in the next section. Below, we discuss the case when triplon quasi-particles become gapless. 
\subsubsection{Ordered Antiferromagnetic Phase}
Suppose, for some values of $\kappa$ and $\zeta$, the dispersion $E_\k$ touches zero at a certain wavevector $\Q$. Then, the corresponding terms in Eqs.~\ref{eq:sbar_cds} and~\ref{eq:lambda} become singular, and there appears a third equation: $E_\Q=0$, relating two unknowns $\sbar^2$ and $\lambda$ in the problem. Following Einstein's approach, we interpret the singularity in the self-consistent equations as Bose-Einstein condensation (BEC) of triplons in mode $\Q$. This gives us a `third' unknown in the form of triplon condensate density $n_c$, which also resolves the problem of closure of the number of unknowns to the number of equations~\cite{rachel}. The physical consequence of triplon condensation, as shown later, is the emergence of non-zero local magnetic moment, giving rise to antiferromagnetic order in the system. 

The closing of the quasi-particle gap fixes $\lambda$ through $E_\Q=0$. The singular terms  with $E_\Q$ as denominator can be suitably expressed in terms of the condensate density, $n_c$. The constraint on the bond-operators implies: $\frac{2}{L}\sum_\k \langle t^\dag_{\k\alpha} t^{ _{ }}_{\k\alpha}\rangle = 1-\sbar^2$. Therefore, the density of condensed triplons in the ground state can be written as: 
\begin{eqnarray}
n_c &=& \frac{2}{L}\langle t^\dag_{\Q\alpha} t^{ _{ }}_{\Q\alpha}\rangle = 1-\sbar^2-
\frac{2}{L}\sum_{\k\neq \Q}\langle t^\dag_{\k\alpha} t^{ _{ }}_{\k\alpha}\rangle \nonumber\\
&=&\frac{5}{2} - \sbar^2 -\frac{3}{L}\sum_{\k\neq\Q}\frac{\lambda-\sbar^2\xi_\k}{E_{\k}} \nonumber \\
&\equiv& \frac{3}{L}\left(\frac{\lambda-\sbar^2\xi_\Q}{E_{\Q}}\right) \label{eq:def_nc} 
\end{eqnarray}
Above, in the last step, Eq.~\ref{eq:sbar_cds} has been invoked.
This is the rule for associating singular terms to $n_c$. If there is more than one $\Q$ for the same condensate, it is understood that we sum over $\Q$ in Eq.~\ref{eq:def_nc}. We now get the following self-consistent equations for $\lambda$, $n_c$ and $\sbar^2$: 
\begin{eqnarray}
\lambda &=& 2\sbar^2\,\xi_\Q \label{eq:lambda_bec}\\
n_c &=& \frac{1}{2\eta_\Q}\left[\lambda-\frac{3\lambda}{L}\sum_{\k\neq \Q}\frac{\eta_\k}{E_\k} -(1-\zeta)J - \frac{8K}{3}\right] \label{eq:nc}\\
\sbar^2&=& \frac{5}{2}-n_c -\frac{3}{L}\sum_{\k\neq \Q} \frac{\lambda-\sbar^2\xi_\k}{E_{\k}} \label{eq:sbar_bec}
\end{eqnarray}
Eq.~\ref{eq:lambda_bec} is a non-trivial solution of $E_\Q=0$ for $\lambda$ (there is a trivial but physically inconsitent solution in $\lambda=0$). Eqs.~\ref{eq:nc} and~\ref{eq:sbar_bec} are the restatements of Eq.~\ref{eq:lambda} and~\ref{eq:def_nc}, respectively. The scheme of iterative calculation in this case goes as follows. For  some initial value of $\sbar^2$, calculate $\lambda$. Feed this $\lambda$ and $\sbar^2$ into Eq.~\ref{eq:nc} to calculate $n_c$. Then, calculate new $\sbar^2$ from Eq.~\ref{eq:sbar_bec}, and close the loop. 

In the BEC ground state, the operators $t^\dag_{\Q\alpha}$, $t^{ _{ }}_{\Q\alpha}$ can be treated as a $c$-number $\tbar$ such that:  $n_c=2\langle t^\dag_{\Q\alpha} t^{ _{ }}_{\Q\alpha}\rangle/L = 6\tbar^2/L$, where $\tbar$ is taken to be real, and independent of  $\alpha$ (because three branches are degenerate). For the spins of a dimer, we can show that: $\langle S_{\R,1\alpha}\rangle = -\langle S_{\R,2\alpha}\rangle=\sqrt{\frac{2}{L}}\,\sbar\,\tbar\,\cos(\Q\cdot\R) = \sbar\sqrt{n_c/3}\,\cos(\Q\cdot\R) $. Since the different components of a spin have same expectation value, we define the local magnetic moments $M_{\R,1}$ and $M_{\R,2}$ as: $\langle S_{\R,1\alpha}\rangle =M_{\R,1}/\sqrt{3}$, and similarly for $M_{\R,2}$. Therefore, $M_{\R,1}=-M_{\R,2}=\sbar\sqrt{n_c}\,\cos(\Q\cdot\R)$. Having a non-zero $n_c$ thus enables the emergence of an AF order with an ordering wavevector $\Q$, and an order parameter (staggered magnetic moment), $M_s=\sbar\sqrt{n_c}$.
\section{Results and Discussion}
\begin{figure}
\centering
\includegraphics[width=7.5cm]{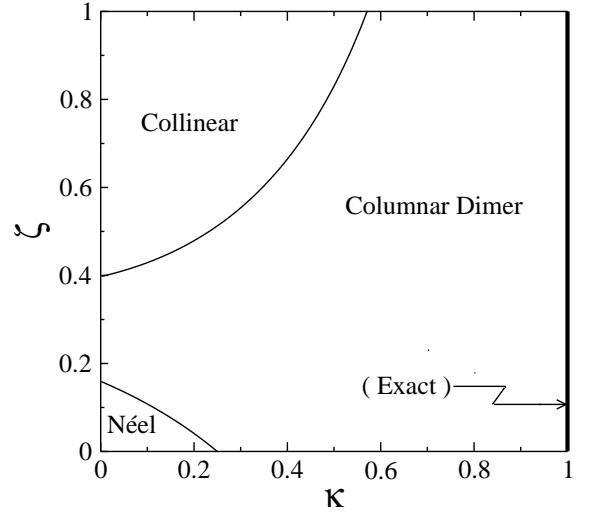}
\caption{The quantum phase diagram of the models $H_1$ and $H_2$ in the triplon mean-field theory. The exact CDS ground state corresponds to $\kappa=1$ line. Also, shown are the regions of N\'eel and collinear ordered AF ground states for $\zeta \lesssim 0.16$ and $\zeta \gtrsim 0.4$, respectively.}
\label{fig:QPD}
\end{figure}

\begin{figure}
\centering
\includegraphics[width=8.5cm]{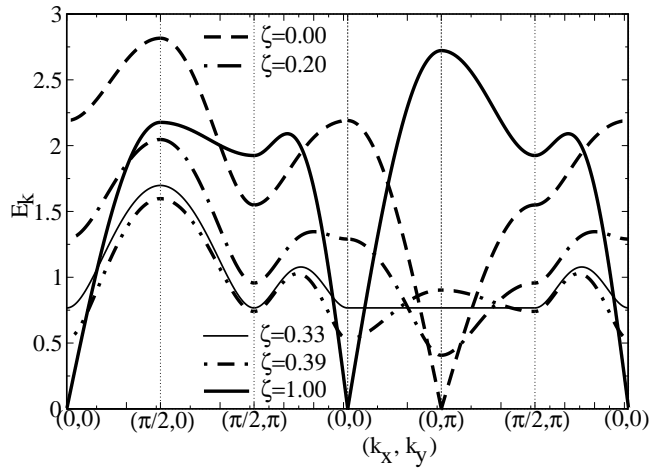}
\caption{The dispersion $E_\k$ of the triplon quasi-particles. Notice that the minimum of each curve is at $\Q_N=(0,\pi)$ for $\zeta<1/3$ and $\Q_C=(0,0)$ for $\zeta>1/3$.}
\label{fig:dispersion}
\end{figure}

\begin{figure}
\centering
\includegraphics[width=4.2cm]{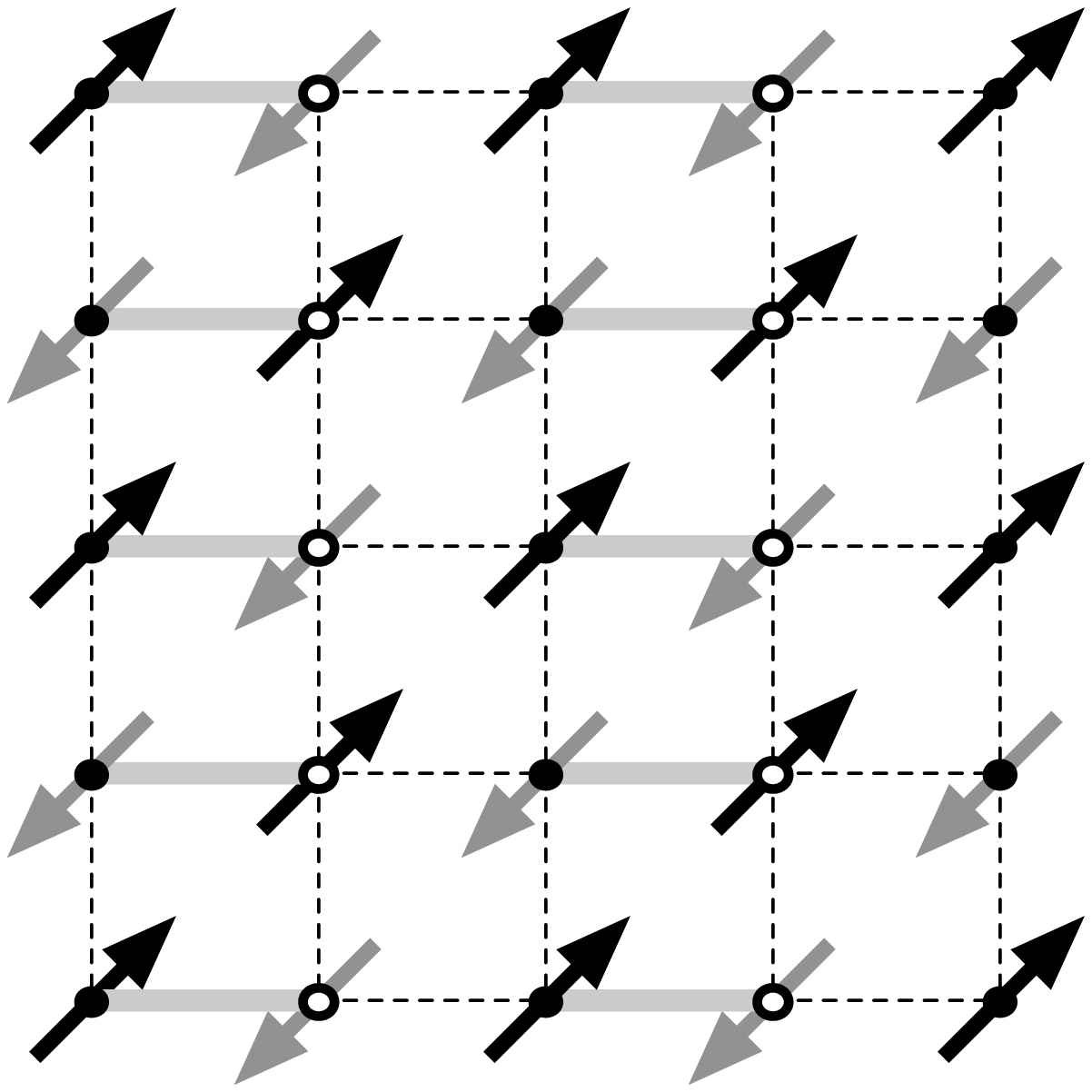}\hfill
\includegraphics[width=4.2cm]{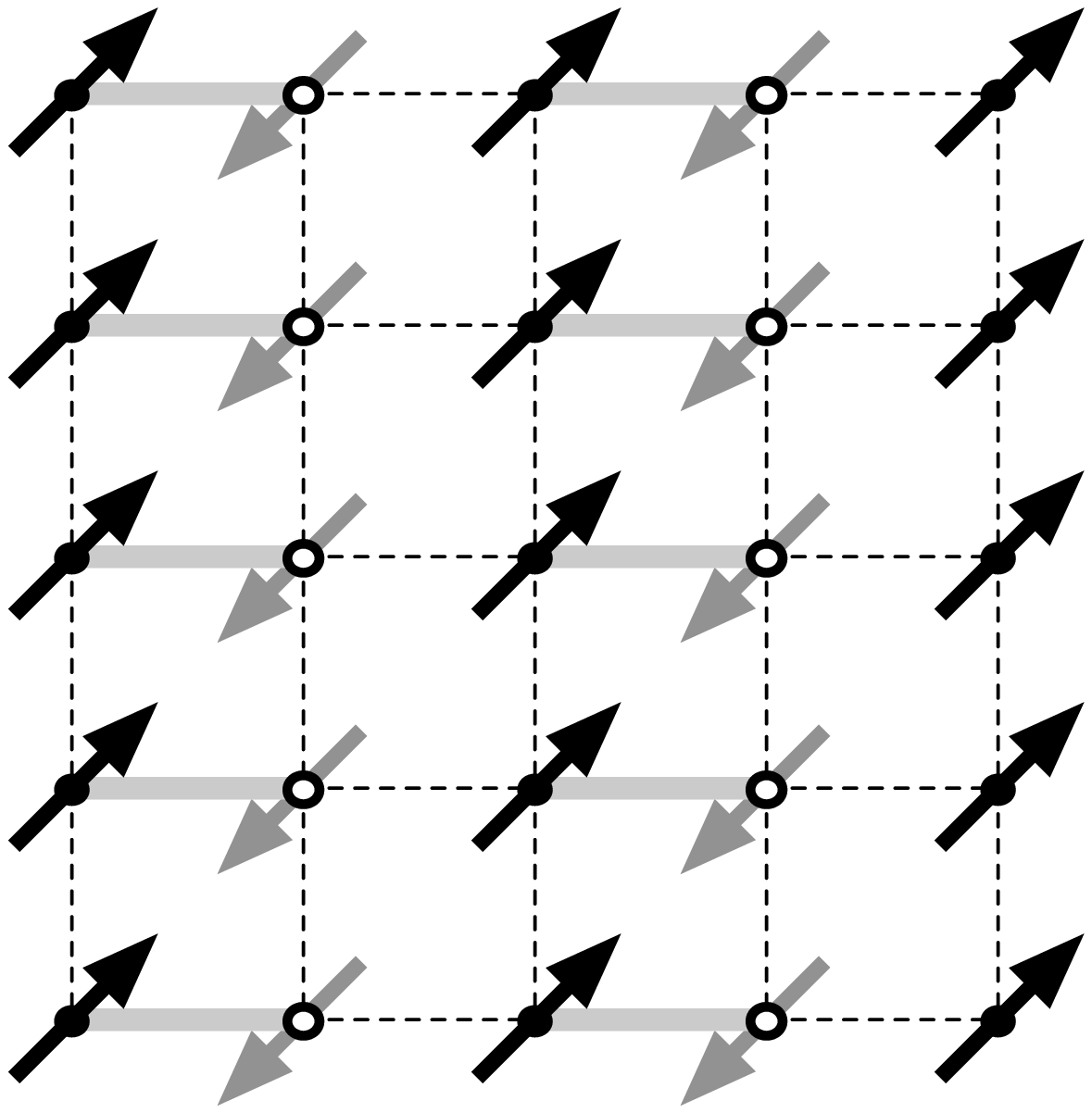}
\caption{Schematic drawings of N\'eel (left) and collinear (right) AF ground states in the triplon mean-field theory. Note that all dimers have identical magnetic moments in the collinear state, hence $\Q_C=(0,0)$. In the N\'eel state, the magnetic moments on neighboring dimers along $y$-direction flip, therefore $\Q_N=(0,\pi)$.}
\label{fig:Neel_collinear}
\end{figure}

We first compute $\sbar^2$ and $\lambda$  in the gapped phase. For $\kappa=1$, we get $\sbar^2=1$, which is same as the {\em exact} result. Also, the flat triplon dispersion with an energy gap of $2.66$ is consistent with our estimate of 8/3 for the gap for a {\em localized} triplon in the exact case. We further compute the mean-field variables for different values of $\kappa$ and $\zeta$. For sufficiently small $\kappa$, our system undergoes a quantum phase transition to ordered AF ground states by closing the triplon gap. The locus of those points in the $\kappa$-$\zeta$ plane, where the triplon gap just vanishes, defines the phase boundary of the gapped CDS phase. Fig.~\ref{fig:QPD} shows the quantum phase diagram  of $H_1$ and $H_2$, as given by the triplon mean-field calculation.
To correctly identify the ordered phases, we must know the minimum of $E_\k$ in the gapped phase. We find two different values of $\Q$, for which the quasi-particle dispersion is minimum (see Fig.~\ref{fig:dispersion}). These are: $\Q \equiv \Q_N=(0,\pi)$ and $\Q_C=(0,0)$ (see Fig.~\ref{fig:CDS_BZ}). The minimum value of the triplon dispersion defines the spin-gap: \( \Delta=\sqrt{\lambda\left( \lambda-2\bar{s}^2\xi_{\bf Q}\right)} \), in the CDS phase. The shift of the minimum of $E_\k$ from $\Q_N$ to $\Q_C$ occurs precisely at $\zeta=1/3$ (which correspond to $J_2/J_1=\tilde{J}_2/\tilde{J}_1=1/2$). The  $\kappa$-$\zeta$ plane has two different regions corresponding to $E_{\Q_N}$ or $E_{\Q_C}$ being zero, in which the ground state has N\'eel or collinear order, respectively. The AF ground states as they emerge on top of the CDS background are pictorially shown in Fig~\ref{fig:Neel_collinear}. This figure also shows how $\Q_N$ and $\Q_C$ correspond respectively to N\'eel and collinear phases. From the gapless side, where $n_c$ (or $M_s$) is the relevant order parameter, the phase boundaries in Fig.~\ref{fig:QPD} are to be understood as those points at which $n_c$ continuously vanishes. 

\begin{figure}
 \centering{%
\includegraphics[width=7.5cm]{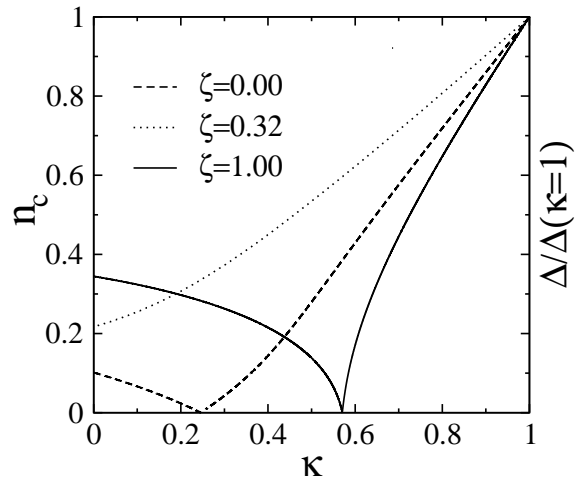}}
\caption{The triplon condensate density $n_c$, and the normalized spin-gap $\Delta/\Delta(\kappa=1)$ versus $\kappa$. $\Delta\neq 0$ gives CDS phase, and $n_c\neq 0$ gives magnetic ordering.}
\label{fig:Gap_nc}
\end{figure}

\begin{figure}
\centerline{\includegraphics[width=7.5cm]{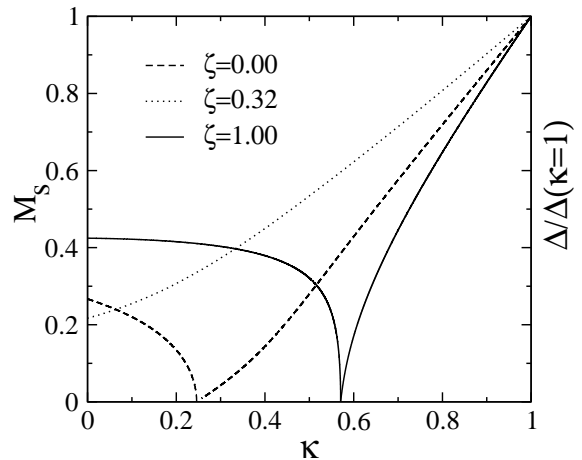}}
\caption{The staggered magnetic moment $M_s$ and normalized spin-gap vs. $\kappa$, for different values of $\zeta$.}
\label{fig:Gap_Ms}
\end{figure}

\begin{figure}
\centering
\includegraphics[width=7cm]{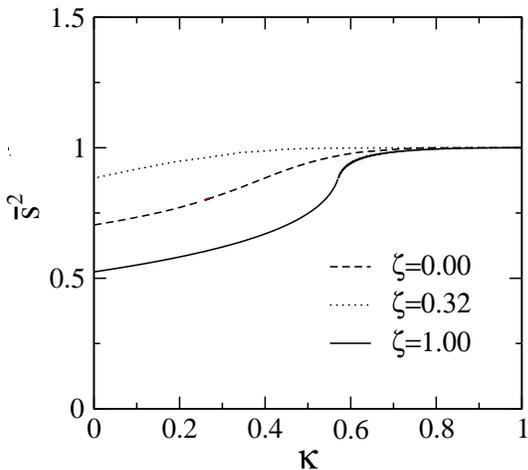}
\caption{The singlet amplitude on a dimer, $\sbar^2$ vs. $\kappa$.}
\label{fig:sbar2}
\end{figure}

\begin{figure}
\centering
\includegraphics[width=7cm]{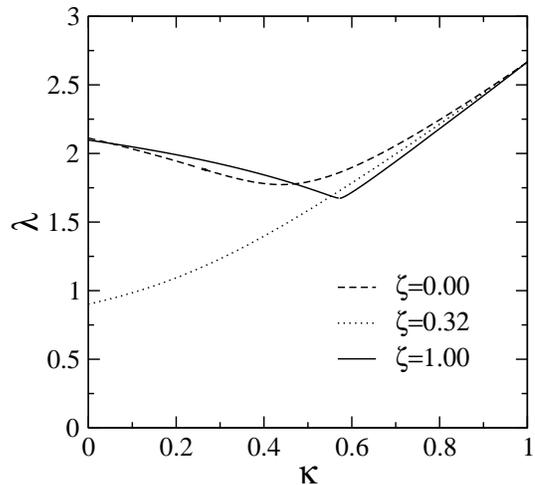}
\caption{Effective chemical potential $\lambda$ vs. $\kappa$. Change in the sign of slope accross the transition suggests a second order transition. There is no such change for $\zeta=0.32$.}
\label{fig:lambda}
\end{figure}

\begin{figure}
\centering
\includegraphics[width=7cm]{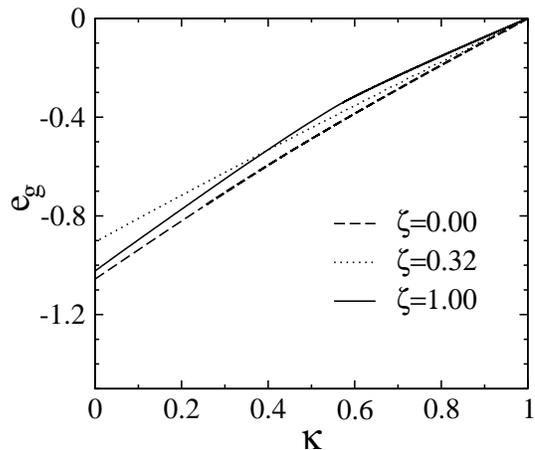}
\caption{Ground state energy per site vs. $\kappa$. }
\label{fig:eg}
\end{figure}

In Fig.~\ref{fig:Gap_nc}, we present $\Delta$ and  $n_c$  as a function of $\kappa$ for different values of $\zeta$. For $\zeta=0$, the gap vanishes at $\kappa\approx 0.25$, and $n_c$ continuously builds up for lower values of $\kappa$. This describes a continuous quantum phase transition from the CDS to N\'eel  phase. For $0\le\zeta\lesssim 0.16$, we always find a $\kappa$, below which the ground state is  N\'eel ordered. For $0.16\lesssim \zeta \lesssim 0.4$, the gapped CDS phase persists for the entire range of $\kappa$. For example, the non-vanishing $\Delta$ for $\zeta=0.32$ in Fig.~\ref{fig:Gap_nc} shows the absence of a quantum phase transition. For $\zeta\gtrsim 0.4$, the CDS phase gives way to collinear AF phase for sufficiently small values of $\kappa$. For $\zeta=1$, the transition to the collinear ordered phase happens at $\kappa\approx 0.57$. The transition to collinear phase in general looks more pronounced as compared to the CDS to N\'eel transition. In Fig.~\ref{fig:Gap_Ms}, we present the same information in terms of the magnetic order parameter, $M_s=\sbar\sqrt{n_c}$.  For completeness, in Figs.~\ref{fig:sbar2},~\ref{fig:lambda} and~\ref{fig:eg}, we also present the data for $\sbar^2$, $\lambda$ and $e_g$ over the whole range of $\kappa$ for a few values of $\zeta$.

We notice that the dimer order of the CDS phase does not wash out completely in the AF ground states. That is, the chosen dimers always have more singlet amplitude as compared to other equivalent bonds (pictorially represented in Fig.~\ref{fig:Neel_collinear}). This is a limitation of the bond-operator mean-field theory, because only {\em one} of the four degenerate states can be chosen to carry out the triplon analysis. Look at the reduction in $\sbar^2$ in the ordered AF phases with respect to its value in the CDS phase in Fig.~\ref{fig:sbar2}. Clearly, this reduction in the N\'eel phase is much less compared to the collinear phase. In the collinear phase at $\kappa=0$, the value of $\sbar^2$ is in fact very close to a crude estimate of 0.5 (for a classical pair of opposite spin-1/2s). This can be qualitatively explained by observing that the collinear phase is also columnar like the CDS state, whereas the N\'eel state has a different magnetic lattice. Therefore, undergoing a transition from CDS to N\'eel phase needs some reorganization, which is unfortunately restricted in the bond-operator mean-field theory. This difference also reflects in the bounds of the intermediate phase for $\kappa=0$, where $J_2/J_1\approx 0.19$ (on the N\'eel side) is considerably smaller compared to the known numerically esitmates~\cite{vbc5}. An alternate approach for treating the degenerate dimer states on equal footing is very much desirable. That will not only improve the results quantitatively, but also help to investigate the possibility of spinons in these models.

The existence of the CDS ground state in $H_1$ and $H_2$, starting from the exact case and extending all the way into the intermediate range $0.16 \lesssim \zeta \lesssim 0.4$ (equivalently, $0.19\lesssim J_2/J_1\lesssim 0.67$) of the pure $J_1$-$J_2$ model, supports the view that there is a spontaneously  dimerized columnar phase between the N\'eel and collinear ordered AF ground states in the $J_1$-$J_2$ model. We believe that the ground state of pure $J_2/J_1=1/2$ model is smoothly connected to the exact CDS ground state of $\kappa=1$ model (like a fixed point Hamiltonian). In other words, an exact numerical study of $H_1$ and $H_2$ is expected to show an adiabatic evolution of the exact CDS ground state to the actual ground state of pure $J_2/J_1=1/2$ model. 
To conclude, the mean-field theory presents a qualitatively robust quantum phase diagram, which may however be improved quantitatively by more exact numerical methods. Also, it reveals a continuous quantum phase transition between the CDS and AF phases, which should further be ascertained by more exact results. In future, we plan to carry out exact numerical studies on these models to address many such questions. Finally, the models $H_1$ and $H_2$ have the necessary ingredients for discussing the possibility of deconfined quantum critical points, and therefore should further be investigated in that direction.

\section{Summary}
We have constructed new models of spin-1/2 quantum antiferromagnets on square lattice, with a fourfold degenerate columnar-dimer ground state. These models consist of a block-interaction part plus the $J_1$-$J_2$ model. The block-interactions either involve two triangles, or a triangle and a square. The interactions in the models are parametrized as $\kappa$ and $\zeta$, such that $\kappa=1$  gives pure block-interactions while $\kappa=0$ is pure $J_1$-$J_2$ model, and $\zeta$ is a measure of $J_2$ with respect to $J_1$. For $\kappa=1$, the ground state is exact, and consists of four CDS configurations.  Away from the exact case, we have formulated a triplon mean-field theory, in terms of the bond-operator bosons, to investigate the domain of stability of the CDS ground state. The CDS phase is characterized by a gap in the triplon dispersion. Since $\kappa$ stabilizes the CDS phase while $\zeta$ (close to 0 or 1) favours antiferromagnetic order, the ground state shows quantum phase transition between the ordered AF and the CDS phases in the $\kappa$-$\zeta$ plane. The onset of an ordered phase is characterized by the closing of the triplon gap. In the gapless phase, the triplons Bose condense, and give rise to an ordered antiferromagnetic ground state, with an ordering wavevector $\Q$ given by those points in the Brillouin zone at which the triplon dispersion touches zero. 
 
The mean-field theory is exact for $\kappa=1$, as it gives a perfect CDS phase. For sufficiently small $\kappa$, two types of ordered AF phases emerge: the N\'eel and the collinear. The N\'eel phase exists in a bounded region close to $(\kappa,\zeta)=(0,0)$, and there is a region of collinear order around $(\kappa,\zeta)=(0,1)$. The rest of the phase diagram is occupied by the stable CDS phase (see Fig.~\ref{fig:QPD}). The triplon gap in the CDS phase continuously goes to zero while approaching the phase boundary. The same is true for staggered magnetization in the ordered AF phase. The triplon mean-field theory thus reveals a picture of various phases and the transitions thereof, in the ground state of these models, which should further be tested against and improved by more elaborate numerical calculations.

\acknowledgments
RK acknowledges the financial support from CSIR. BK acknowledges the financial support under the project No. SR/FTP/PS-06/2006 from the DST (India).
\bibliography{manuscript_4CDS}

\begin{thebibliography}{44}
\bibitem[{fn-({\natexlab{a}})}]{fn-1}
\bibinfo{note}{This is estimated by putting a 1/2 for each spin in $H_0$ and
  $V_{1,~2}$. Also, in the exact CDS state, it is easy to check that $\langle
  V_{1,~2}\rangle =0$, and $\langle H_0\rangle/L = -K$, which is consistent
  with our simplistic estimate.}
\bibitem[{fn-({\natexlab{b}})}]{fn-2}
\bibinfo{note}{In principle, the exact ground state in the rectangular case,
  $L_x\ne L_y$, also has four CDS configurations. But now, the $x$ and $y$ type
  CDS states are not related through $\calC_4$}.
\bibitem[{fn-({\natexlab{c}})}]{fn-3}
\bibinfo{note}{If both the blocks of an interaction term are satisfied, then
  the fraction of a dimer on this block-interaction term will be more than 1/8.
  Hence, it will be less than 1/8 for some other block-interaction term, which
  will be left unsatisfied}.
\bibitem[{fn-({\natexlab{d}})}]{fn-4}
\bibinfo{note}{At the simplest level, this dependence already shows up in the
  Majumdar-Ghosh case (first-neighbor coordination=2). This is just a way of
  appreciating the degeneracies predicted in Ref.~\cite{ReadSach_NPhyB}.}
\end{thebibliography}

\end{document}